\documentstyle[aps,prb,epsfig,float,twocolumn]{revtex}
\newcommand{\be}{\begin{equation}}
\newcommand{\ee}{\end{equation}}
\newcommand{\bea}{\begin{eqnarray}}
\newcommand{\eea}{\end{eqnarray}}
\newcommand{\rd}{\mbox{d}}
\newcommand{\p}{\partial}

\begin{document}
\twocolumn[\hsize\textwidth\columnwidth\hsize\csname
@twocolumnfalse\endcsname

\title{Interactions Suppress Quasiparticle Tunneling at Hall Bar Constrictions}

\author{Emiliano Papa$^{1}$ and A.H. MacDonald$^{1,2}$}

\address{$^1$Department of Physics, The University of Texas, Austin TX 78712}
\address{$^2$California Institute of Technology, Pasadena CA 91125}

\date{\today}

\maketitle

\begin{abstract}

Tunneling of fractionally charged quasiparticles across a 
two-dimensional electron system on a fractional quantum Hall plateau is expected to 
be strongly enhanced at low temperatures.
This theoretical prediction is at odds with recent experimental studies
of samples with weakly-pinched quantum-point-contact constrictions, 
in which the opposite behavior is observed.  We argue here that 
this unexpected finding is a consequence of electron-electron interactions near 
the point contact.
 

\end{abstract}
\vskip2pc]

\begin{figure}
\unitlength=1mm
\begin{picture}(70,35)
\put(5,3){\epsfig{file=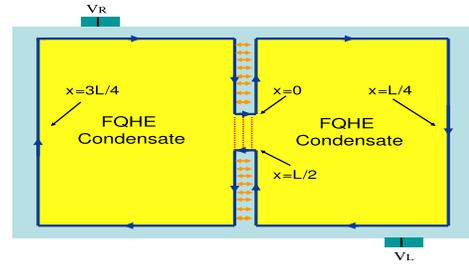,angle=90,angle=90,angle=90,height=38mm,width=66mm}}
\end{picture}
\caption{Schematic illustration of a Hall bar with a split-gate constriction.
Samples similar to this have been used to study tunneling of fractionally charged quasiparticles across an
incompressible two-dimensional electron gas region with filling factor $\nu=1/3$.  For weakly pinched gates 
the $\nu=1/3$ plateau extends through the point-contact.  Our 
model measures position clockwise along the edge starting from the top of the split gate at $x=0$, so
that the bottom of the split gate is at $x=L/2$ where $L$ is the total edge perimeter and positions are 
understood to be defined modulo $L$.  Quasiparticle tunneling from top to 
bottom across the point contact is equivalent to electron backscattering by
the constriction.  The quasiparticle tunneling current $I^{qp}_{Tunn}$ leads to a voltage drop 
$V = h I^{qp}_{Tunn}/e^2\nu$ across the constriction 
and to an identical deviation from perfect quantization for Hall voltages 
measured on either left or right hand sides of the Hall bar.}
\label{fig:one}
\end{figure}

\noindent
{\em Introduction:}
One-dimensional fermion systems have attracted enduring interest because they are converted 
from Fermi liquids to Luttinger liquids (LL)\cite{LuttingerLiquid} by interactions.
Convincing experimental evidence of LL behavior has been discovered in 
a number of quasi-one-dimensional systems,
including carbon nanotubes\cite{NanotubeExpt}, semiconductor quantum wires\cite{WireExpt},
and the edges\cite{QHEdgeFirstExpt,QHEdgeReview} of incompressible two-dimensional electron systems
on quantum Hall plateaus.  From a theoretical point of view, quantum Hall edge systems 
are especially interesting\cite{QHEdgeFirstTheory} because they do not reduce to Fermi-liquids even when 
interactions between charge fluctuations along the edge are neglected.
Although edge systems can be complex,\cite{QHEdgeTheory,Reconstruction}
those that surround the simplest 
fractional quantum Hall state at $\nu=1/3$ appear\cite{QHEdgeFirstExpt,TunnelingDOS} 
to be reasonably well described
by the simplest consistent model which has a single chiral edge channel.
A key prediction\cite{BackScatteringTheory}
of theory
is that tunneling of fractionally charged quasiparticles across a  
constriction in a two-dimensional electron gas, like the 
one illustrated schematically in Fig.~\ref{fig:one}, is universally enhanced at small 
bias voltages and low temperatures.
This Letter is motivated by recent experiments\cite{RoddaroHeiblum,Otherexplanation} in which the opposite behavior 
is consistently observed.  The discrepancy exists even though theory appears 
to describes other non-trivial edge-related properties correctly, 
including the decrease in current noise 
due to the fractional quasiparticle charge\cite{FracCharge} and the suppressed tunneling 
density-of-states.\cite{TunnelingDOS}
In this Letter we argue that these surprising observations are a consequence of 
interactions near the constriction.


In order to achieve tunneling of fractionally charged quasiparticles it is necessary to bring 
opposite edges of a Hall bar into proximity by placing a split gate over the top.
A typical sample has 
an overall length and width $\sim 100\mu$m, and a gate width which is $\sim 100$ times smaller.  
Simply put, our idea is that the edge states 
on the left and right sides of the split gate should be 
regarded as the counter-propagating chiral 
channels of a one-dimensional fermion system defined on the constriction, rather than as the
left and right portions of the top or bottom chiral channels of the overall Hall bar.  
As we show, interactions across the split-gate suppress quasiparticle tunneling across the point-contact 
formed by its opening.  We argue that these interactions can be sufficiently strong to make weak quasiparticle
tunneling an irrelevant perturbation and produce the low-bias tunneling {\em suppression} seen in many experiments.
We elaborate on this idea by studying a simple toy model that captures essential features of the experimental geometry.

\noindent{\em The model:}
We consider a Hall bar with a constriction whose edge 
encloses a singly connected area in which an incompressible 
quantum Hall state with filling factor $\nu$ has been established.
(See Fig.~[\ref{fig:one}].) 
Low energy physics on a quantum Hall plateau\cite{QHEdgeFirstTheory}
may be described in terms of an edge density collective coordinate $\rho(x)$.  The effective Hamiltonian 
\begin{equation} 
H = \frac{1}{2}\int_{0}^L \rd x \int_{0}^L \rd x' \rho(x) U(x,x') \rho(x') \quad ,
\label{eq:hamiltonian}
\end{equation}
where  $U(x,x') =   {\delta^2   E[\rho(x)]} / {\delta \rho(x)\delta \rho(x')}  \left.\right|_0$. 
%
$U(x,x')$ can be expressed as a sum of microscopic exchange and confining potential
interactions that are expected\cite{ZuelickeEnergy} to cancel approximately, and 
the Coulomb energy given approximately by 
$U(x,x') \approx e^2/{[\epsilon |\vec{r}(x)-\vec{r}(x')|]}$
where $\vec{r}(x)$ is a two-dimensional coordinate at the edge. 
Following arguments first articulated by Wen,\cite{QHEdgeFirstTheory}
it follows from the fractional quantum Hall effect that the single-chiral-channel
model is quantized by imposing the commutation relations 
\bea
[\rho(x),\rho(x')]=-(i\nu/2\pi)\p_x\delta(x-x') \quad .
\label{comm_relat}
\eea
The quasiparticle tunneling operator is expressed below in terms of   
the chiral boson field $\phi(x)$, related to the density (in our convention) by 
$\rho(x) = \nu \p_x\phi(x)/2\pi$.

With these approximations the quadratic edge action is completely specified by the sample
geometry allowing any density-fluctuation correlation function to be evaluated numerically.  
Our argument is most simply explained,  
however, by considering a simple toy model for which the relevant correlation functions
can be evaluated analytically.  We consider the Hamiltonian $H=H_0+H_1+H_2$ where 
\bea
H_0 &=& \pi v_F \hspace{-1mm} \int_{0}^L \hspace{-2mm} \rd x \rho(x)\rho(x)\; , \;
H_1 = g_1 \pi v_F \hspace{-1mm} \int_{0}^{L} \hspace{-2mm} \rd x \rho(x) \rho(L-x)\; , \nonumber \\
&& \hspace{8mm} H_2 = g_2 \pi v_F \int_{0}^{L} \rd x \rho(x) \rho(L/2-x). 
\label{Hammodel} 
\eea
In Eq.~(\ref{Hammodel}) $g_1$ and $g_2$ are dimensionless interaction parameters,
$H_0$ accounts for interactions between nearby points along the edge,
$H_1$ for interactions across the split-gate, and $H_2$ for interactions between the top and the bottom
of the Hall bar.  For a Hall bar with 
a constriction we expect that $g_1$ is close to but smaller than $1$ and that $g_2$ is close to $0$.
The parameter
which characterizes the strength of local interactions, $v_F$, is the edge wave velocity 
for $\nu=1$ and $g_1=g_2=0$.  
The action for this model, 
\bea
S &=& \hspace{-0.1mm} \frac{1}{L}\hspace{-0.1mm} \sum_{i>0}\hspace{-0.1mm} \int_{\tau} 
\hspace{-0.1mm}\Big[ 
 \phi^*(q_i,\tau)\frac{i\nu q_i }{2\pi} \partial_{\tau} \phi(q_i,\tau) +
 \frac{\nu^2 q_i^2 v_F}{2\pi} \Big[ \left|\phi(q_i,\tau)\right|^2  
\Bigr.\nonumber \\
&-&
\Bigl.\bigl.
\left(g_1+(-)^{i}g_2\right)
 \Bigl(\phi^*(q_i,\tau) \phi^*(q_i,\tau)+ {\rm c.c.} 
\Bigr) \Bigr]
\Bigr] \,,
\eea
has normal modes with eigenenergies and operator formalism creation operators given by 
$E_{n\pm} = \hbar v_F q_n \nu \left[1-g_{\pm}^2\right]^{1/2}$
and $A_n=\cosh(\theta_{\pm}) \; a_n + \sinh(\theta_{\pm}) \; a_n^{\dagger}$.
Here $a_n$ and $a_n^{\dagger}$ are edge wave creation and annihilation operators for $g_i=0$ which are 
proportional to the $ q = \pm 2 \pi n/L$ Fourier components of the edge density,  
the plus signs apply for even $n$, the minus signs for odd $n$,
$\tanh(2\theta_{\pm})= g_{\pm}$ and $g_{\pm}=g_1 \pm g_2$. 

\noindent
{\em Quasiparticle Tunneling I-V Characteristics:} 
%
The operator which creates a quasiparticle of charge $\nu$ at the 
edge of the system is\cite{QHEdgeFirstTheory}
${\hat \psi}_{qp}(x) = e^{i \nu \phi(x)}/(2\pi)^{1/2}$.
The operator for quasiparticle tunneling from top to bottom across the constriction 
is therefore 
\bea
&&T_{\rm qp}(0,\frac{L}{2};t) 
=\frac{1}{2\pi}  \exp\bigl\{i \nu \phi(\frac{L}{2},t)\bigr\}\exp\left\{-i \nu \phi(0,t)\right\}  =
\nonumber \\[2mm]
&&\exp\Bigl[- i2e^{\theta_{-}} 
\sum_{n>0}^{ \rm odd} \sqrt{\frac{\nu}{n}} \left[A_n e^{-i E_n t/\hbar} - A_n^{\dagger} e^{i E_n t/\hbar} 
\right]\Bigr] \,.
\label{qptunneling}
\eea
Note that this operator depends only on the interaction combination $g_1-g_2$; 
repulsive interactions across the constriction play the same role as attractive interactions 
across the Hall Bar.  

The quasiparticle tunneling current can be evaluated by treating the tunneling term in the 
Hamiltonian, $H_{\rm Tunn} = (\Gamma T_{qp} + \Gamma^* T_{\rm qp}^{\dagger})$, perturbatively.  At 
leading order this gives the usual Fermi Golden rule result: 
{\small
\begin{equation}
I^{\rm qp}_{\rm Tunn} = \left[1-\exp\left(-\frac{eV}{k_BT}\right)\right]  \frac{e^*}{\hbar} 
\int_{-\infty}^{+\infty} \rd t e ^{ieVt} \;  
G^{\rm qp}_{\rm Tunn}(t) \,, 
\label{I_Tunn}
\end{equation} }
where the quasiparticle tunneling correlation function is 
\bea
\nonumber
&&G^{\rm qp}_{\rm Tunn}(t) = \left< T_{\rm qp}^{\dagger}(t)  T_{\rm qp}(0) \right>
=\frac{|\Gamma|^2}{(2\pi)^2}
\exp\Bigl[ -4\nu e^{2\theta_{-}} \sum_{n>0}^{ \rm odd} \frac{1}{n} 
\Bigl[
\bigr. \Bigr.
\\ \Bigl. \bigl. &&
\bigl(n_B(E_n)+1\bigr)\left(1-e^{-\frac{iE_nt}{\hbar}}\right) 
+ n_B(E_n)\left(1-e^{\frac{iE_nt}{\hbar}}\right)\Bigr]\Bigr]  \,,
\label{I_Tunn_sum}
\eea
and $n_B(E)$ is the Bose distribution function.  For $k_B T > \hbar v_F /L$ the sum can be replaced by 
an integral and we find (up to a constant factor) that 
\bea
G^{\rm qp}_{\rm Tunn}(t)= \frac{|\Gamma|^2}{(2\pi)^2} \left(i \frac{v_{-} \beta}{\pi}  \sinh \frac{\pi t}{\beta} \right)^{-2\nu e^{2 \theta_{-}}} \quad.
\eea
(For finite system sizes the quasiparticle tunneling correlation function takes the form
\bea
G^{\rm qp}_{\rm Tunn}(t)= \frac{|\Gamma|^2}{(2\pi)^2} F^{-2\nu e^{2\theta_-}}(z_-)
\tilde{F}^{2\nu e^{2\theta_-}}(z_-) \;,
\eea
where $z_-=v_-t$ and  $F$ and $\tilde{F}$ are Euler elliptic
$\vartheta$-function ratios: $F(z) = \vartheta_1({\pi z}/{L}| i{v\beta}/{L})/
{\vartheta_1(-i{\pi \alpha}/{L}| i{v\beta}/{L})}$, and
$\tilde{F}(z)=\vartheta_2({\pi z}/{L}|i{v\beta}/{L})/{\vartheta_2(-i{\pi \alpha}/{L}| i{v\beta}/{L})}$
and $\alpha$ is an infinitesimal used to regularize the $n$-summation.)

Fourier transforming the large $L$  correlation function yields 
the final expression for the tunneling current: 
{\small
\begin{equation}
I^{\rm qp}_{\rm Tunn}  =  \frac{2 e^* |\Gamma|^2 \sin(\pi d)}{\pi v_- h} 
\left(\frac{v_-\beta}{2\pi} \right )^{1-2d} \hspace{-4mm} {\rm Im}
\left\{ B(d-i\frac{eV\beta}{2\pi},1-2d) \right\}
,
\label{qp_current}
\end{equation}}
where $B$ is the Euler beta function. 
The key quantity in this expression is $d = \nu e^{2\theta_{-}}$ which is the scaling dimension of 
the tunneling operator.

The significance of $d$ in the quasiparticle tunneling current expression is more apparent
in the simpler expressions that apply in low and high temperature 
limits.  For $eV\beta/2\pi \ll 1$  
{\small
\bea
I^{\rm qp}_{\rm Tunn}  =
&&\frac{ e^* |\Gamma|^2}{v_- h} \left(\frac{v_-\beta}{2\pi} \right )^{1-2d} \; 
\hspace{-2mm}
\frac{\Gamma^2(d)}{\Gamma(2d)}\cdot
\nonumber\\[2mm]
&&\hspace{-10mm}\cdot \frac{eV\beta}{2\pi}\left[1+\left(\frac{\pi^2}{6}-\zeta(2,d)\right)
\left(\frac{eV\beta}{2\pi}\right)^2 +\cdots\right]
\quad,
\eea}
where $\zeta\left(2,d\right)$ is the Riemann
zeta function. 
Note that $\zeta\left(2,d\right)$ is a monotonically decreasing function of $d$ and equals 
$\pi^2/6$ for $d=1$.   It follows that for $d<1$ the conductance $dI^{\rm qp}/dV$ diverges with $T$
like $T^{2d-2}$ in the Ohmic region, and decreases like $T^{2d-4}V^2$ at higher voltages.
In the opposite case when $d>1$ the low bias conductance goes to zero like 
$T^{2d-2}$ in the Ohmic region, and increases like $T^{2d-4}V^2$ at higher voltages.
%
In the high bias limit, $eV\beta/2\pi \gg 1$,
\begin{equation}
I^{\rm qp}_{\rm Tunn}  = \frac{ e^* |\Gamma|^2}{h v^{2d}_- }
\frac{\quad (eV)^{2d-1}}{\Gamma(2d)}
\quad ,
\end{equation}
so that the conductance varies as $V^{2d-2}$.   
For non-interacting electrons on the integer Hall edge $\nu=1$, $g_1=g_2=0$, $d=1$, and the tunneling conductance 
is independent of both bias voltage and temperature.  The chiral LL
prediction\cite{BackScatteringTheory} that fractional quasiparticle tunneling increases at low 
temperatures follows
from the factor of $\nu$ in the expression for $d$ which makes $d <1$ and backscattering relevant.
The key observation here is that interactions can alter this prediction if $e^{2\theta_{-}}$ 
is sufficiently large.

\noindent
{\em Interaction Parameter and Scaling Dimension Estimates:}
To determine whether or not interactions can alter the relevance of quasiparticle tunneling we 
need to estimate 
$\exp\{2\theta_{-}\} = [(1+g_{-})/(1-g_{-})]^{1/2}$.
Because the microscopic Coulomb interactions are long ranged the effective 
interaction strengths depends weakly (logarithmically) on wavevector. 
The local interaction parameters in our model, $v_F$, is the Coulomb interaction 
Fourier transform cut-off at short distances by $w_{\rm edge}$, the width of the region 
near the edge over which the charge density falls from
its bulk value to zero.   It follows that 
$v_F(k) \sim  -e^2 \ln(kw_{\rm edge})/ \epsilon\pi\hbar$.  
Non-local interactions across the 
constriction and across the Hall bar are given by similar expressions, except that the short distance
cut-offs are respectively the width of the constriction and the width of the Hall bar:  
$v_F(k) g_1 \sim -{e^2}\ln(kw_{\rm sg})/\epsilon\pi\hbar$
where $w_{\rm sg} \sim 0.3 {\rm \mu m}$ is the width of the split gate, and 
$v_F(k) g_2 \sim  0$ except for the case of long thin Hall bars. 
For quantitative estimates it is important to realize that 
$\epsilon$ 
should be taken as the mean\cite{wassermeier} of the dielectric 
constant of the (GaAs) host semiconductor and vaccuum.
The wavevector $k =2\pi n/L $ that is relevant to the experimental behavior is determined by the condition 
$E_{n-} \sim k_B T$, which defines a thermal length $L_{T}$ with a value $\sim 100 {\rm \mu m}$ for 
temperatures $\sim 0.1$ Kelvin, in the middle of the typical measuring range.  Since this length is not 
substantially larger than the length of the split gate, quasiparticle tunneling should be sensitive only to 
interactions along the gate and our idealized model is realistic.  It follows that 
$g_1 \sim \ln(L_T/w_{\rm sg})/[\ln(L_T/w_{\rm sg})+\ln(L_T/w_{\rm edge})] \sim 0.84$,
and that $\exp(2\theta_{-}) \sim 3.4$, larger than the value required to transform quasiparticle tunneling 

\noindent 
{\em The closed constriction limit:} 
Similar calculations can be performed for electron tunneling between 
the disconnected regions on the left and right hand sides of the Hall bar
that enclose FQH liquids with filling fractions $\nu_L$ and $\nu_R$ respectively. 
We find that 
the {\em electron} tunneling-tunneling correlation function is
\bea
\label{tunn_electr}
&&G^{\rm el}_{\rm Tunn} (0,\frac{L}{2};t-t') =
\frac{|\Gamma|^2}{(2\pi)^2} \cdot
\nonumber \\[0mm]\nonumber
&&\cdot \left(i \frac{v_L\beta}{\pi}\sinh\frac{\pi (t-t')}{\beta}\right)^{-\frac{1}{ \nu_L }\left(\cosh 2\theta-\sqrt{\frac{\nu_L}{\nu_R} }\sinh 2\theta   \right) } 
\cdot
\nonumber\\[0mm] &&
\cdot \left(i \frac{v_R\beta}{\pi}\sinh\frac{\pi (t-t')}{\beta}\right)^{-\frac{1}{ \nu_R }\left(\cosh 2\theta-\sqrt{\frac{\nu_R}{\nu_L} }\sinh 2\theta   \right) } 
\label{closed_constr}
\quad.
\eea
When $\nu_1=\nu_2\equiv \nu$ and for large system sizes we find that 
the scaling dimension is $d^{\rm el} = \exp\{-2\theta_{-}\}/\nu$. 
Since $d \,d^{\rm el} = 1$, the top-bottom quasiparticle 
tunneling process for an open constriction is relevant whenever left-right 
{\em electron} tunneling through a closed constriction is irrelevant and vice-versa.
 
\noindent
{\em Discussion:} 
The model we study here has a number of interesting symmetries that are captured by the 
general quasiparticle-quasiparticle correlation function
{\small
\bea
\label{greens_funct}
&&\left<R_{\rm qp}^\dagger(x,t) R_{\rm qp}(x',t')\right>=
\\[1.5mm]
&& \hspace{1.5mm}\left[F^{-\frac{\nu}{2}\cosh^2\theta_+}(z_+ - z_+')F^{-\frac{\nu}{2}\cosh^2\theta_-}(z_- - z_-')\right] \cdot
\nonumber
\\[1.5mm]\nonumber
&& \cdot \left[F^{-\frac{\nu}{2}\sinh^2\theta_+}(\bar{z}_+ - \bar{z}_+')F^{-\frac{\nu}{2}\sinh^2\theta_-}(\bar{z}_- - \bar{z}_-')\right] \cdot
\nonumber\\[1.5mm]\nonumber
&&\cdot \left[\frac{\tilde{F}^{-\frac{\nu}{2}\cosh^2\theta_+}(z_+ - z_+')
              \tilde{F}^{-\frac{\nu}{2}\sinh^2\theta_+}(\bar{z}_+ - \bar{z}_+')
}{            \tilde{F}^{-\frac{\nu}{2}\cosh^2\theta_-}(z_- - z_-')
              \tilde{F}^{-\frac{\nu}{2}\sinh^2\theta_-}(\bar{z}_- - \bar{z}_-')
}\right] \cdot
\\[1.5mm]\nonumber
&&
\cdot \left[ \frac{F^{-\frac{\nu}{2}\sinh\theta_+ \cosh\theta_+}(z_++\bar{z}_+')F^{-\frac{\nu}{2}\sinh\theta_- \cosh\theta_-}(z_-+\bar{z}_-')}
{            F^{-\frac{\nu}{2}\sinh\theta_+ \cosh\theta_+}(z_++\bar{z}_+) F^{-\frac{\nu}{2}\sinh\theta_- \cosh\theta_-}(z_-+\bar{z}_-)  }\right] \cdot
\\[1.5mm]\nonumber
&&
\cdot \left[ \frac{F^{-\frac{\nu}{2}\sinh\theta_+ \cosh\theta_+}(\bar{z}_+ + z_+')F^{-\frac{\nu}{2}\sinh\theta_- \cosh\theta_-}(\bar{z}_- +z_-')}
{            F^{-\frac{\nu}{2}\sinh\theta_+ \cosh\theta_+}(\bar{z}_+'+ z_+') F^{-\frac{\nu}{2}\sinh\theta_- \cosh\theta_-}(\bar{z}_-'+ z_-')  }\right] \cdot
\\[1.5mm]
&&
\cdot \left[\frac{\tilde{F}^{-\frac{\nu}{2}\sinh\theta_+ \cosh\theta_+}(z_+       +  \bar{z}_+')
      \tilde{F}^{-\frac{\nu}{2}\sinh\theta_+ \cosh\theta_+}(\bar{z}_+ +  z_+')
}{    \tilde{F}^{-\frac{\nu}{2}\sinh\theta_- \cosh\theta_-}(z_-+\bar{z}_-')
      \tilde{F}^{-\frac{\nu}{2}\sinh\theta_- \cosh\theta_-}(\bar{z}_- + z_-')}
\right] \cdot
\nonumber
\\[1.5mm]\nonumber
&&\cdot \left[
\frac{\tilde{F}^{-\frac{\nu}{2}\sinh\theta_- \cosh\theta_-}(z_-       +  \bar{z}_-)
      \tilde{F}^{-\frac{\nu}{2}\sinh\theta_- \cosh\theta_-}(z_-' +  \bar{z}_-')
}{    \tilde{F}^{-\frac{\nu}{2}\sinh\theta_+ \cosh\theta_+}(z_++\bar{z}_+)
      \tilde{F}^{-\frac{\nu}{2}\sinh\theta_+ \cosh\theta_+}(z_+' + \bar{z}_+')}
\right]
\quad ,
\eea
}
where $z_-=x-v_-t$, $z_+=x - v_+t$.
In these expressions $F$ and $\tilde{F}$ are again ratios of Euler's elliptic $\vartheta$-functions.
The model system is invariant under reflection at either $x=0$ (the horizontal axis) or 
at $x=L/2$ (the vertical axis).  The scaling 
dimensions of a tunneling process depends in general 
on the edge points that are linked.  
For example one can see that the quasiparticle-quasiparticle 
correlation function and the scaling dimensions of the tunneling process remain unchanged under the shift
$z,z' \rightarrow z+L/2, z'+L/2$, 
but the roles of $g_1$ and $g_2$ are interchanged by the shift $z,z' \rightarrow z+L/4, z'+L/4$.  
We have 
concentrated here on tunneling across a horizontal 
Hall bar with a constriction defined by a vertical gate with an opening at its center.
This general 
expression suggests that quasiparticle tunneling properties can be altered in interesting ways by 
changing the sample geometry.   
One example is that the static correlation function scaling dimension 
for $0\simeq x'\ll x \ll L$ and $g_1=g_2$, 
$d = \nu (4 + K+3/K)/8 $ where $K=\exp(-2\theta_{+})$,
compared to the standard LL expression $d= ({\nu}/{2})(K+{1}/{K})$. 
for the standard LL.

The edges of fractional incompressible quantum Hall states are unique 
in that they maintain non-Fermi liquid behavior independent of the  
details of interactions along the edge.  Their non-Fermi liquid behavior arises most 
fundamentally from  the appearance of the fractional filling factor $\nu$ in the 
density-fluctuation commutators.  For this reason it is usually expected that 
edge state properties, for example the scaling dimension of quasiparticle tunneling,
should be universal.  Experiments nevertheless find find that quasiparticle tunneling is 
non-universal and often irrelevant, as noted by Rosenow and Halperin\cite{Otherexplanation}.
In this paper we have proposed that this behavior arises from 
the character of 
interactions near quantum point contacts created by narrow split gates.
We have shown that 
interactions between edge waves approaching and leaving the quantum point contact, tend to suppress 
quasiparticle tunneling.  Our numerical estimates, based on the geometry of a sample studied in a 
recent paper by Roddaro et al.\cite{RoddaroHeiblum}, suggest that this explanation is plausible.
Since the effect requires that the split gate be narrow over a long distance,
it should be possible to test this explanation experimentally.  We emphasize that our quasiparticle-tunneling 
theory applies only in the case of an open point contact, in which the $\nu=1/3$ incompressible fractional Hall
regime is continuous from one side of the point contact to the other, and our electron-tunneling theory only
to strongly pinched contacts.  We expect that non-universal interaction effects will also play a role 
in the interesting intermediate regime\cite{Roddaro2004}
in which incompressible regions with smaller filler factors may be formed 
inside the point contact.  Finally, we remark that non-universal interaction effects will also play a 
role in samples with a simple long-thin Hall bar geometries, if the temperature can be reduced well below the 
energy of an edge wave with a wavelength equal to the {\em width} of the Hall bar.  For example,
for Hall bars with a width of $\sim 1 {\rm \mu m}$, non-universal effects due to $g_2$ interactions
across the width of the Hall bar should start to become important below $\sim 10 K$ on a $\nu=1$ plateau.  

This work was supported by the National Science Foundation under grant DMR 0115947 and 
by the Welch Foundation.  We acknowledge helpful interactions
with Moty Heiblum, Vittorio Pellegrini, Stephano Roddaro, Roberto Raimondi,
Giovanni Vignale, and Alexei Tsvelik.  AHM acknowledges support from the Moore Scholar program and the 
hospitality of the California Institute of Technology.

\end{document}